\documentclass[showpacs,nofootinbib,showkeywords,amssymb,amsmath,floatfix,prd]{revtex4-1}
\usepackage[a4paper, total={6.5in, 10in}]{geometry}
\usepackage{CJKutf8}

\usepackage{graphicx,color,amsmath,amsxtra}
\usepackage{epsf}
\usepackage{amssymb}
\usepackage{enumerate}
\usepackage{hhline}
\usepackage{array}
\usepackage{tabularx}
\usepackage[unicode]{hyperref}
\usepackage{float,subfig}
\usepackage{caption}
\usepackage{tikz}
\usetikzlibrary{decorations.pathmorphing,patterns}

\usepackage{csquotes}

\usepackage{graphicx}% Include figure files
\usepackage{dcolumn}% eqnarray table columns on decimal point

\usepackage{tikz}
\begin{document}
\begin{CJK}{UTF8}{ipxm}
%\begin{CJK}{UTF8}{bsmi}
%\setlength{\parskip}{5pt}
%\renewcommand{\bibname}{References}

\newcommand{\W}{{\vec W}}
\newcommand{\n}{{\hat n}}
\newcommand{\hn}{{\hat n}}
\newcommand{\hr}{{\hat r}}
\newcommand{\hD}{{\hat D}}
\newcommand{\bea}{\begin{eqnarray}}
\newcommand{\eea}{\end{eqnarray}}
\newcommand{\A}{{\vec A}}
\newcommand{\valpha}{{\vec \alpha}}
\newcommand{\vbeta}{{\vec \beta}}
\newcommand{\D}{{\hat D}}
\newcommand{\nn}{\nonumber}
\newcommand{\pro}{\partial}
\newcommand{\Int}{\displaystyle{\int}}
\newcommand{\mn}{{\mu\nu}}

%\ead{$^*$gore@mail.nctu.edu.tw}
%\date{\today }
\date{March 6, 2021}

\title{ A new look at the SU$_2$ gauge monopole}
\author{W. F. Kao\footnote{homegore09@nycu.edu.tw}}
\address{{}
Institute of Physics, National Yang-Ming Chiao-Tung University, Hsin Chu, Taiwan
}
\begin{abstract}
%\begin{abstract}
An $SU_2\times U_1$ scalar vector model with a scalar doublet $\varphi$ is reviewed for the study of possible magnetic monopole solution. An eigenvalue equation $\hat n^a  \sigma^a \varphi_\pm =\pm \varphi_\pm$ is shown to induce a set of monopole solutions specified by the unit vector $\hat n$. 
It is shown clearly that monopole solution has to do with the eigenvalue equation and an unique covariant combination of the non-abelian gauge field. 
It is also shown that a new set of monopole solutions is presented as a generalization of the monopole solutions known as Cho-Maison monopole. We also show that the EM field and field tensor defined by CM solution is effectively the same as the effective covariant field tensor introduced in this paper. Possible implication is also discussed in the literature.
%\end{abstract}
\end{abstract}


\maketitle

\section{Introduction}
The properties of magnetic monopole has been a focus of research interest ever since Dirac \cite{1} proposed the possible existence of monopole in 1931.
Non-abelian $SO_3$ gauge monopole was then proposed by 't Hooft in 1981 \cite{2,3,4}. 
In particular, the non-abelian monopoles found by Wu and Yang \cite{3,prl80}  shows that the pure $SU_2$ gauge theory admits a point-like monopole.
't Hooft  and Polyakov \cite{4,prasad} have then constructed a finite energy monopole solution in Georgi-Glashow model. 
Similar monopole solution for the grand unified model has also been found in Ref. \cite{dokos}.

The gauge field model is given by
\begin{eqnarray} \label{action}
{\cal L}=-\frac{1}{4} {\rm Tr}~  G^2 -\frac{1}{4}    Y^2-\frac{1}{2}(\nabla\phi)^{\dagger}\nabla\phi-V(\phi) 
\end{eqnarray}
with a complex scalar doublet $\phi$ coupled to the $SU_2$ gauge field $A_\mu$.
Here the scalar field potential is given by 
\begin{eqnarray}
V=-\frac{\lambda}{8}(\phi^{\dagger}\phi-v^2)^2 ~. 
\end{eqnarray}
with a constant parameter $v$ signifying the vacuum state of the scalar potential.
In addition, the covariant derivative of the scalar field is defined as 
$ \nabla\phi = (\partial-iA)\phi$ for $SU_2$ (or $\nabla \phi=  (\partial-iA -iY/2)\phi$ for $SU_2 \times U_1$) model with $A=A^{a}T^a$ and $T^a=\sigma^a/2$. 
Here the space-time subindex is not shown explicitly for convenience.
Note that $\sigma^a$ is the Pauli matrix and $\tilde G_{\mu \nu} =G_{\mu \nu}+Y_{\mu \nu}/2= i [~ \nabla_\mu, \nabla_\nu ~]= \partial_\mu A_\nu-\partial_\nu A_\mu -i [~A_\mu, A_\nu~] +Y_{\mu \nu}/2$ represents the gauge field tensor of $SU_2 \times U_1$ gauge fields. Here $G_{\mu \nu}= \partial_\mu A_\nu-\partial_\nu A_\mu -i [~A_\mu, A_\nu~] $ and $Y_{\mu \nu}= \partial_\mu Y_\nu-\partial_\nu Y_\mu$.

In addition, the  $2 \times 2$ matrix $T^a=\sigma^a /2$ denotes the generator of the gauge algebra $su_2$ with $\sigma^a$  obeying the relations
\begin{eqnarray}
& \sigma^a \sigma^b=\delta^{ab} +i \epsilon^{abc}  \sigma^c ~,
\\ & [~T^a, T^b~] = i \epsilon^{abc} T^c 
~.
\end{eqnarray}

Note that the gauge group $SO(3,R)$ is isomorphic to $SU(2,C)$ with similar algebraic structure $su_2 \simeq so_3$.
Hence the real $SO_3$ gauge invariant model can be written as
\begin{eqnarray}
{\cal L}=-\frac{1}{4}{\rm Tr}~  G^2-\frac{1}{2}(\nabla\phi)^t\nabla\phi-V(\phi) 
\end{eqnarray}
with a real scalar triplet $\phi$ coupled to the $SO_3$ gauge field $A_\mu$.

A new type of non-abelian gauge field dyon solutions known as Cho-Maison (CM) monopole are found in Ref. \cite{plb97,yang}.
Related references can also be found in Ref. [10-52]. 
An  $SU_2$  scalar field ansatz proposed to take the form 
$$\varphi_c^t=i( \cos (\theta/2)  \exp(-i \varphi), - \sin (\theta/2)~)$$
is the key to the monopole solution.
Here $\hat r =(\sin \theta \cos \varphi, \sin \theta \sin \varphi, \cos \theta)$ with $\theta(x), \varphi(x)$ denoting the spherical angles associated with the unit radius vector $\hat r$.
The scalar doublet ansatz is in fact a special solution to the eigenvalue equation $\hat n \varphi_c \equiv  \hat n^i \cdot  \sigma^i  \varphi_c = -  \varphi_c$ with $\hat n =\hat r$. \cite{CK1,CK2}
 In addition, we also use the same notation $\hat n =\hat n \cdot  \sigma$ to represent a dressed $2 \times 2$ traceless and hermitian matrix  with the property $\hat n \hat n =I$.
 Matrix $\hat n$ and 3-vector $\hat n$ can be read off directly from the equation without any confusion.
The unit vector can be generalized to $\hat n =(\sin \theta_1 \cos \varphi_1, \sin \theta_1 \sin \varphi_1, \cos \theta_1)$ with $\theta_1(x), \varphi_1(x)$ denoting the gauged spherical angles (functions of space time) associated with the unit vector $\hat n(x)$.
As a result, a new set of monopole solutions can be generated automatically.
For simplicity the gauged angles will be restricted to be functions of spherical angles in this paper.

The scalar field ansatz $\varphi^t =(\hat r^1, \hat r^2, \hat r^3)$ proposed, in Ref. \cite{3}, to induce the $SO_3$ monopole  is also the eigenvector  to the eigenvalue equation $\hat n^a \cdot  T^a \varphi_0 = 0$  with  $\lambda =0$ and $\hat n =\hat r$.
Here the $3 \times 3$ matrix $T^a$ represents the generators of the gauge group $SO_3$.
In particular, the existence of monopole solution has to do with the identification of a special combination of gauge field that exhibits the monopole structure with \cite{CK1}
\begin{eqnarray}\label{4.28}\label{mono}
  F_{\mu\nu} 
&&  = \partial_\mu B_\nu - \partial_\nu B_\mu - \epsilon^{abc}n^a \partial_\mu n^b \partial_\nu n^c 
\end{eqnarray}
The monopole term $\epsilon^{abc}n^a \partial_\mu n^b \partial_\nu n^c $ can also be shown to be uniquely given by the identification of gauge field tensor \cite{CK1}
\begin{eqnarray}\label{F1}
 F_{\mu\nu} && \equiv \lambda  \varphi_\lambda^\dagger G_{\mu\nu} \varphi_\lambda + i { \lambda  } \; \left  [ \; ( \nabla_\mu \varphi_\lambda)^\dagger  \nabla_\nu \varphi_\lambda - \; (\nabla _\nu \varphi_\lambda)^\dagger  \nabla_\mu \varphi_\lambda\; \right ] ~
\end{eqnarray}
with the help of the eigenvalue equation $\hat n \varphi_\pm = \pm \varphi_\pm$.
Here the summation over all  repeated $\lambda=\pm 1$ is not written explicitly for convenience.
Indeed, the monopole structure $\epsilon^{abc}n^a \partial_\mu n^b \partial_\nu n^c$ with an unit vector $\hat n$ can only be related to the scalar field consistently with the  gauge covariant eigenvalue equation $\hat n^a T^a \varphi_\lambda = \lambda \varphi_\lambda$, or equivalently $\hat n \varphi_\lambda = \lambda \varphi_\lambda$,  for doublet scalar field with 3 properly chosen generators $T^a$ forming a closed subgroup. 

Hence it was pointed out in 1983 that the existence of gauge monopole solution has to do with the identification of gauge field $\phi$ with the eigenvector to the eigenvalue equation $\hat n^i T^i \varphi_\lambda= \lambda \varphi_\lambda$, or equivalently, \cite{CK1}
\begin{eqnarray}\label{eigen}
\hat n  \varphi_\lambda= \lambda \varphi_\lambda 
\end{eqnarray}
with a difference in the eigenvalue $\lambda$ by a factor of 2.
Here the scalar field $\phi$ can be decomposed as $\phi=\chi \varphi_0 \varphi_\lambda$ with $\varphi_\lambda$  denotes the normalized eigenvector of the above eigenvalue equation with eigenvalue $\lambda=\pm 1$. 
$\chi$ and $\varphi_0= \exp [~i \psi~]$ represent the norm and abelian phase factor of $\phi$ respectively.
Note that the norm $\chi$ and abelian phase factor $\varphi_0$ of the doublet field $\phi$ will not affect the eigenvector as a solution to the eigenvalue equation above.
In particular, the CM monopole can be generalized straightforwardly by setting $\hat r \to \hat n$.
Details of the generalized ansatz will be also presented in this paper.
We will also show explicitly that the field tensor $F$ of CM monopole also agrees with the form given by \eqref{F1}.

Alternatively, tracking the gauge field in its original matrix form will lead us to the intrinsic nature hidden in the geometric properties of the scalar doublets.
Hence an outline of Ref. \cite{CK1} in matrix formulation will be presented in this paper for reference.
A general discussion of the properties and identities associated with the eigenvector will also be presented in this paper.
The materials shown in this paper will hopefully be helpful in generalizing any known monopole solutions in a more systematic way. \cite{cho}

This paper will be organized as follows.
A brief introduction to the monopole structure is presented in Se. I. 
In Sec. II, physical meaning of the eigenvalue equation will be presented.
Useful properties of the eigenvector will be shown in Sec. III.
We will also show how to prove that the unique, covariant effective field tensor defined in Eq. \eqref{F1} leads to
the monopole structure shown in Eq. \eqref{mono} with the help of the eigenvalue equation \eqref{eigen}.
This proof will be presented in Sec. IV. 
A set of generalized CM monopole solutions will be shown in Sec. V. 
We will also show that the EM field and field tensor defined by CM solution is effectively the same as the effective covariant field tensor introduced in this paper. 
 
A conclusion and discussion will be drawn in the section of conclusion.
Some useful review and proof are also presented in the appendices for completeness.
 
\section{physical meaning of the eigenvalue equation}
The existence of gauge monopole has to do with the topological structure of the coupled scalar field.
As mentioned above, the  the scalar field $\phi=\chi \varphi$ can be decomposed as a norm, or length scale, and a normalized scalar field with the property $\varphi^\dagger \varphi=1$. 
Due to the scale symmetry, derived from the dimensionless gauge Lagrangian $-G^2/4$, the $\chi$ field does not couple directly to the gauge field directly. 
Hence the physical behavior of the gauge field is directly related to the topological structure of the normalized $\varphi$. 
As a result, a nontrivial second homotopy group $\Pi_2(M)$, with $M$ the geometry of the scalar field solution, is known to be critical to the existence of the monopole.
Indeed, the topology of the $SO_3$ real normalized triplet scalar field $\varphi$ can thus be shown to be $\varphi^a \varphi^a=1$ as an $S^2$.
The second homotopy group of $SO_3$ scalar field can be shown to be $\Pi_2(\varphi)=\Pi_2(S^2)={\bf Z}$.
Hence the real $SO_3$ theory is possible to admit a monopole solution.

The geometry of the normalized $SU_2$ scalar field $\varphi$ can be shown to be $S^3$.
Indeed, we can parametrized $\varphi^t = (n_0+in_3, n_+)$ with $n_\pm = n_1 \pm i n_2$.
Hence the normalized condition $\varphi^\dagger \varphi=1$  gives $n_0^2+n_1^2+n_2^2+n_3^2=1$.
As a result, it is equivalent to the geometry of $S^3$.
Hence it is known the $SU_2$ gauge theory does not support the existence of monopole due to the fact that $\Pi_2(\varphi)=\Pi_2(S^3)=0$.

Indeed, in order to have a monopole solution, we have to find a combination of gauge field as a reduced $U_1$ gauge field  with a covariant, effective field tensor takes the following form
\begin{eqnarray}
 F_{\mu\nu} = \partial_\mu B_\nu- \partial_\nu B_\mu  -\epsilon^{abc} \partial_\mu \hat  n^b \partial_\nu \hat n^b~.
\end{eqnarray}
Here the last term, known as monopole term, is the key to the existence of the monopole solution.
%In addition, the unit real  3-vector ${\hat n}^i(x)$, ${\hat n}^i {\hat n}^i=1$, can be any arbitrary function of space time.
In order to construct a special combination of gauge field tensor with the above property, we need to connect the unit vector ${\hat n}^a$ with the the normalized scalar field $\varphi$.
The only consistent way to do is to appeal to the eigenvalue equation 
\begin{eqnarray}
\hat n   \varphi_\lambda=\lambda \varphi_\lambda .
\end{eqnarray}
Here $\varphi_\lambda$ is the normalized eigenvector of the above eigenvalue equation with eigenvector $\lambda=\pm 1$. 
This eigenvalue equation thus imposes a constraint on the normalized scalar field up to an $U_1$ scalar field $\varphi_0=\exp [i \psi]$.
Indeed, we can show that the eigenvector of above equation gives
\begin{eqnarray}
& \varphi_+^t =\left (\cos {\theta_1 \over 2}  \exp [~ - i \varphi_1 ~], ~  \sin {\theta_1 \over 2} \right ) ~,
\\
&  \varphi_-^t =\left (\sin {\theta_1 \over 2}  \exp [~ - i \varphi_1 ~], ~ - \cos {\theta_1 \over 2} \right )~.
\end{eqnarray}
Note that the CM monopole doublet $\varphi_c= i \varphi_-$ is still an eigenvector of $\hat n$ with $\hat n \varphi_c=- \varphi_c$.
This is because that an additional phase factor or scale factor will not affect the eigenvalue equation. 
Indeed, in addition to the scale factor $\chi$, there is a factor $\varphi_0 =\exp [ i \psi ]$ undetermined by the eigenvalue equation.
In fact, the gauge transformation associated with the $SU_2$ gauge field $A^u=UAU^{-1}+i U \partial U^{-1}$, with $U= \exp [~ i  \alpha ^i (x) T^i ~]\in SU_2$, has much to do with the  phase transformation $\varphi_0$.
Extending the $SU_2$ model by introducing an additional $U_1$ gauge field $Y_\mu$, the role played by the phase field $\varphi_0$ will become more transparent.
Indeed, a change in the $U_1$ phase scalar $\varphi_0$ is to be taken care of by the gauge transformation of the the $U_1$ gauge field $Y_\mu$ via the covariant derivative
\begin{eqnarray}
\tilde \nabla \varphi = (\partial -iA-iY/2)\varphi ~.
\end{eqnarray}
By writing the normalized scalar field as $\varphi=\varphi_0 \varphi_2$, we can show that the covariant derivative decouples immediately as
\begin{eqnarray}
\tilde \nabla \varphi =\varphi_0  (\partial -iA)\varphi_2 + \varphi_2  (\partial -iY/2)\varphi_0 ~.
\end{eqnarray} 
Here $\varphi_2$ retains only the $S^2$ geometric part of the gauge field configuration.
In fact, by virtue of the symmetry, $\phi=\chi \varphi_0 \varphi_2$, a more general form of the covariant derivative should include the Weyl vector meson $S_\mu$ by defining
\begin{eqnarray}
\tilde \nabla \phi = (\partial -S -iA-iY/2)\phi ~.
\end{eqnarray}
As a result, the Weyl vector meson transforms as $S'=S-\partial \lambda$ if $\chi'=\Lambda \chi$.
In addition, the gauge fields transform as $A'=U_2AU_2^{-1}+i U_2 \partial U_2^{-1}$ and $Y'=U_0YU_0^{-1}+2i U_0 \partial U_0^{-1}$ with $U_2= \exp [~ i  \hat \alpha ^a (x) T^a ~] \in SU_2$ and $U_0=  \exp [~ i  \alpha(x) ~] \in U_1$.
Accordingly, the covariant derivative breaks into
\begin{eqnarray}
\tilde \nabla \phi =\varphi (\partial -S) \chi + \chi \varphi_0  (\partial -iA)\varphi_2 + \chi \varphi_2  (\partial -iY/2)\varphi_0 ~.
\end{eqnarray} 
In addition, the kinetic term also decoupled in-part as
\begin{eqnarray}
-{1 \over 2} \tilde \nabla \phi ^\dagger \tilde \nabla \phi =
-{1 \over 2}  \nabla \chi  \nabla \chi
-{1 \over 2}\chi^2   \nabla \varphi_0 ^\dagger  \nabla \varphi_0
-{1 \over 2} \chi^2  \nabla \varphi_2 ^\dagger  \nabla \varphi_2 
+  \chi^2 \varphi_0^\dagger  \nabla \varphi_0 \varphi_2^\dagger  \nabla \varphi_2 ~.
\end{eqnarray} 
Here $ \nabla \chi = (\partial -S )\chi$,   $\nabla \varphi_0 = (\partial - iY/2)\varphi_0$ and  $ \nabla \varphi_2 = (\partial-iA)\varphi_2$ represent the associated covariant derivatives.

In summary, the scalar field $\phi = \chi \varphi_0 \varphi_2$ can be decomposed as three completely different fields that exhibit scale transformation, abelian and $SU_2$ non-abelian gauge transformations associated with the Weyl vector meson $S_\mu$, abelian gauge field $Y_\mu$  and non-abelian gauge field $A_\mu^i$.
This is also the reason that a complete Higgs doublet model invites the presence of the $U_1$ field in addition to the $SU_2$ field in a natural way.
For convenience, we will write $\varphi_2$ as $\varphi$, the normalized $SU_2$ scalar doublet, associated with the $SU_2$
gauge transformation unless $\varphi_0$ needs to be specified apparently.

Note that the eigenvectors $\varphi_\pm$ spans the normalized doublet space $S^2$.
Hence any relevant physical field in this space can be expanded into two independent components with the help of the base vector $\varphi_\pm$.   
Indeed, $\varphi= \alpha_+ \varphi_+ +\alpha_- \varphi_-$ for any normalized doublet field with coefficients $\alpha_i$ satisfying $|\alpha_+|^2 +|\alpha_-|^2=1$.
As shown above, $\varphi_2$ has only two degrees of freedom left, forming the $S^2$ geometry, once the $U_1$ sector $\varphi_0$ is removed.
Hence the eigenvector is nothing but a reparametrization of the scalar field parameters. 

In addition, the gauge field tensor $G_{\mu\nu}$ can undergo spontaneously symmetry breaking by settling down to the vacuum eigenstate specified by $\chi=v$.  
Here the gauge field tensor $G_{\mu\nu}$ is defined as 
\begin{eqnarray}
G_{\mu \nu}^a = \partial_\mu A_\nu^a- \partial_\nu A_\mu^a + \epsilon^{abc} A_\mu^b A_\nu^c
\end{eqnarray}
or equivalently  
\begin{eqnarray}
G_{\mu \nu}=G_{\mu \nu}^a T^a= \partial_\mu A_\nu- \partial_\nu A_\mu - i[ \; A_\mu,\; A_\nu \;]
\end{eqnarray}
with $A_\mu = A_\mu^a T^a$.
The generators of the gauge group satisfy the relations 
$ [T^a , T^b]= i \epsilon^{abc} T^c$, $< T^aT^b> =k \delta ^{ab}$ 
with $k= 1/2$ for 2D and $k=2$ for 3D representation.
Generalization the results from $SU_2$ models to $SO_3$ model is straightforaward.
Hence we will focus on the $SU_2$ models from now on.

Note that the eigenvectors $\varphi_\pm$ form an orthonormal basis of the 2D $S^2$ space.
In order to find the effective gauge tensor with a monopole term\cite{CK1}
\begin{eqnarray}\label{Fab}
 F_{\mu\nu} = \partial_\mu B_\nu- \partial_\nu B_\mu  -\epsilon^{abc} \partial_\mu \hat  n^b \partial_\nu \hat n^b
\end{eqnarray}
we can show that the only consistent way is to define the effective $U_1$ field $B_\mu = \hat n^a  A_\mu^a$ and the effective $U_1$ gauge field tensor $F$ as \cite{CK1}
\begin{eqnarray}\label{Fmunu}
 F_{\mu\nu} && \equiv \lambda  \varphi_\lambda^\dagger G_{\mu\nu} \varphi_\lambda + i { \lambda } \; \left  [ \; ( \nabla_\mu \varphi_\lambda)^\dagger  \nabla_\nu \varphi_\lambda - \; (\nabla _\nu \varphi_\lambda)^\dagger  \nabla_\mu \varphi_\lambda\; \right ] ~.
\end{eqnarray}
Here the summation over all $\lambda=\pm 1$ is not written explicitly for convenience.
A proof will be presented shortly in Sec. IV.

The monopole is induced by the mapping of an unit vector $\hat n(x)$ to the vacuum scalar field through the eigenvalue equation.
Hence the eigenvector plays an important role in defining the effective gauge field for the existence of monopole solution.

\section{some useful properties of the eigenvector}
As shown earlier that the eigenvectors $\varphi_\pm$ spans the normalized doublet space $S^2$.
Hence any relevant physical field $\varphi$ in this space can be expanded into two independent components as $\varphi= \alpha_+ \varphi_+ +\alpha_- \varphi_-$ with coefficients $\alpha_i$ satisfying $|\alpha_+|^2 +|\alpha_-|^2=1$.
In addition, some useful identities can be shown and listed as below
\begin{eqnarray}\label{eq22}
&& \hat n^a = {\lambda \over 2} \varphi_\lambda^\dagger \sigma^a \varphi_\lambda
\\
&& \varphi_\pm ^\dagger \sigma^a \varphi_\pm  = \pm \hat n^a
\\
&&  \varphi_\lambda \varphi_\lambda ^\dagger= I
\\
&& 
\varphi_\delta^\dagger \partial_\mu \hat n \varphi_\lambda = (\lambda - \delta)\varphi_\delta^\dagger \partial_\mu \varphi_\lambda 
\end{eqnarray}
Note that repeated index is understood to be summed over all $\lambda$ and $\delta$.

Note that the unit vector $\hat n=\hat n^a \sigma^a$ satisfies the relation $\hat n \hat n = {\bf I}$.
Hence the gauge fields $A(\hat n, \partial \hat n)$ as a function of $\hat n$ and $\partial \hat n$ can be expanded to the leading orders as
\begin{eqnarray}
&&A_\mu =  f_\mu \hat n +i f_2 \hat n \partial_\mu \hat n 
\end{eqnarray}
with parameters $f_\mu$ and $f_2$ if the gauge fields solutions, as function of the unit vector $\hat n$, have to do with the eigenvalue equation $\hat n \varphi_\pm =\pm \varphi_\pm$. 
This is also because the eigenvector in fact spans the $S^2$ space.
In particular, the CM monopole solution assumed the gauge field solutions with $f_\mu =A(r)\partial_\mu t$ and $f_2=f(r)-1$. 
Note that the CM type gauge field is assumed to be
\begin{eqnarray}
Y_\mu=h_\mu -(1 - \cos \theta_1) \partial_\mu \varphi_1 
\end{eqnarray}
with $h_\mu=B(r) \partial_\mu t$. 
In addition, CM monopole also assumes that $\hat n =\hat r$. 
Hence $\theta_1=\theta$ and $\varphi_1=\varphi$. \cite{plb97}

%As a result, for the solution $\nabla \varphi_+=0$, we can thus show that
%\begin{eqnarray}
%&& S_\mu =h_\mu -f_\mu - { 1 - \cos \theta_1 \over 2} \partial_\mu \varphi_1 ~.
%\end{eqnarray}
%with an additional parameter $h_\mu$.
%This is exactly the extension of the CM monopole ansatz with an arbitrary unit vector $\hat n$.

Note also that the eigenvectors can be shown to be related to the unit vector $a_+^t=(1,0)$ and $a_-^t=(0, 1)$  through the gauge transformation
\begin{eqnarray}
&& \varphi_+= i \exp [~-i \varphi_1~] U^{-1} a_+ = i \exp [~-i \varphi_1~] \exp \left  [~ -{i \over 2} \theta_1 \hat m ~ \right ]  a_+ ,
\\
&& \varphi_c=U^{-1} a_- =- i  \exp  \left [~ -{ i \over 2} \theta_1 \hat m ~ \right ]  a_-
\end{eqnarray}
with $\hat m= (\sin \varphi_1, - \cos \varphi_1,0) $ as an unit vector normal to $\hat n$, i.e. $\hat m \cdot \hat n =0$. 
In addition, the notation $\hat m =\hat m^i  \sigma^i$ is also used to represent a $2 \times 2$ hermitian traceless matrix for convenience.
This also reflects the fact that the eigenvector is nothing but a special reparametrization of the field parameters of $\varphi$ in $S^2$ up to an $U_1$ phase factor.

Note also that the dyon charge conjugation operator $M=- i \sigma^2$ can thus be shown to be the solution to the dyon charge conjugation transformation\cite{CK1}
\begin{eqnarray}
&& \varphi^c=M \varphi ^*,
\\
&& \nabla \varphi^c =M (\nabla \varphi)^*
\\
&& M T^* M^{-1}=-T
\end{eqnarray}
such that $M \tilde G_{\mu \nu}^* M^{-1}=-\tilde G_{\mu \nu}$ \cite{CK1} with dyon charges change sign as a result of the dyon charge conjugation transformation.
Hence, we can show that the solution to $M$ is $M=- i \sigma^2$ unique up to an arbitrary constant phase factor.
In addition we can also show that $MM^* =-1$. 
It can also be shown that $\varphi_\pm$ are in fact charge conjugate to each other
\begin{eqnarray} 
&& M \varphi_+ ^* =\varphi_-,
\\
&& M \varphi_-^* = - \varphi_+
\end{eqnarray}

\section{monopole term and the eigenvector}

We will show in this section how to prove that the monopole term is present in the combination of the effective gauge field tensor defined by
\begin{eqnarray}
 F_{\mu\nu} && \equiv \lambda  \varphi_\lambda^\dagger G_{\mu\nu} \varphi_\lambda + i { \lambda } \; \left  [ \; ( \nabla_\mu \varphi_\lambda)^\dagger  \nabla_\nu \varphi_\lambda - \; (\nabla _\nu \varphi_\lambda)^\dagger  \nabla_\mu \varphi_\lambda\; \right ] ~
\end{eqnarray}
with $\varphi_\lambda$ the eigenvector.
Here the projection of $G_{\mu\nu}$ is defined by
\begin{eqnarray}
  \hat{n}^a G_{\mu\nu}^a 
 &=&
 \lambda\varphi_\lambda^\dagger G_{\mu\nu}\varphi_\lambda=  \hat{n}^a( \partial_\mu A_\nu^a - \partial_\nu A_\mu^a + e\epsilon^{abc}A_\mu^b A_\nu^c ) 
    \end{eqnarray}
    with $\hat{n}^a = (\lambda/2) \varphi_\lambda^\dagger \sigma^a \varphi_\lambda$, $B_\mu=  \hat{n}^a A_\mu ^a=  \lambda\varphi_\lambda^\dagger A_{\mu}\varphi_\lambda$. 
    Here $A_\mu= A_\mu ^a \sigma^a/2 \in SU_2$.
Note that the coupling constant $g$ is not written explicitly for convenience.
They can restored straightforwardly.
Indeed we will show that
\begin{eqnarray}
  F_{\mu\nu} 
  &=& \partial_\mu B_\nu - \partial_\nu B_\mu - \epsilon^{abc}\hat{n}^a \partial_\mu \hat{n}^b \partial_\nu \hat{n}^c .
    \end{eqnarray}
In fact, the projected gauge field tensor can be rearranged as
\begin{eqnarray}
  \hat{n}^a G_{\mu\nu}^a 
   &=& \partial_\mu B_\nu - \partial_\nu B_\mu + \mathbb{X} _{\mu\nu}+ \mathbb{Y}_{\mu\nu}
  \end{eqnarray}
with the field tensor $\mathbb{X}_{\mu\nu}$  and $\mathbb{Y}_{\mu\nu}$  defined as
  \begin{eqnarray}
\mathbb{X}_{\mu\nu} 
&=& \epsilon^{abc}\hat{n}^a A_\mu^b A_\nu^c = -i \lambda \varphi_\lambda^\dagger \left[ A_\mu, A_\nu \right]\varphi_\lambda 
\\ 
\mathbb{Y} _{\mu\nu}
&=& A_\mu^a \partial_\nu \hat{n}^a - A_\nu^a \partial_\mu \hat{n}^a.
=
\lambda\left[ A_\mu^a \partial_\nu(\varphi_\lambda^\dagger T^a \varphi_\lambda ) - (\mu \leftrightarrow \nu) \right]
\end{eqnarray}
% The eigenvector equation adopted in the appendix, following the notation used in Ref. \cite{CK1} is
% \begin{eqnarray}
% \hat n \cdot T \varphi_\lambda = \lambda \varphi_\lambda
% \end{eqnarray}
% with $\lambda=\pm 1/2$ and $T^a=\sigma^a /2$ the $SU_2$ generators.
In addition, it can be shown that
 \begin{eqnarray}
\mathbb{X}_{\mu\nu} 
&=& 
\lambda\left[-i(D_\mu \varphi_\lambda)^\dagger D_\nu \varphi_\lambda +i(\partial_\mu \varphi_\lambda)^\dagger D_\nu \varphi_\lambda - \varphi_\lambda^\dagger A_\mu \partial_\nu \varphi_\lambda \right] - (\mu \leftrightarrow \nu)  
\\ 
\mathbb{Y} _{\mu\nu}
&=& \lambda\left[i(\partial_\mu \varphi_\lambda)^\dagger \partial_\nu \varphi_\lambda -i(\partial_\mu \varphi_\lambda)^\dagger D_\nu \varphi_\lambda + \varphi_\lambda^\dagger A_\mu \partial_\nu \varphi_\lambda \right] - (\mu \leftrightarrow \nu)
\end{eqnarray}
Note again that the summation over all $\lambda$ is not written explicitly for convenience.
As a result, the combination of the $\mathbb{X}_{\mu\nu} $ and $\mathbb{Y}_{\mu\nu}$ reduces to the result 
\begin{eqnarray}
 \mathbb{X}_{\mu\nu} +\mathbb{Y}_{\mu\nu} =i \lambda \left[ (\partial_\mu \varphi_\lambda)^\dagger \partial_\nu \varphi_\lambda
 -
 (\nabla_\mu \varphi_\lambda)^\dagger \nabla_\nu \varphi_\lambda - (\mu\leftrightarrow\nu) \right]
\end{eqnarray}
Hence the combination gives
\begin{eqnarray}
 Z_{\mu\nu}=  \mathbb{X}_{\mu\nu} +\mathbb{Y}_{\mu\nu} + i \lambda \left[ 
 (\nabla_\mu \varphi_\lambda)^\dagger \nabla_\nu \varphi_\lambda - (\mu\leftrightarrow\nu) \right] =i \lambda \left[ (\partial_\mu \varphi_\lambda)^\dagger \partial_\nu \varphi_\lambda - (\mu\leftrightarrow\nu)
 \right].
\end{eqnarray}
In addition, it can be shown first that
\begin{eqnarray} \label{C1}
\epsilon^{abc}\hat{n}^a \partial_\mu \hat{n}^b \partial_\nu \hat{n}^c 
  &=& -i \frac{\lambda}{4}\left[ (\lambda - \delta)^2 
  (\partial_\mu \varphi_\lambda)^\dagger \varphi_\delta \varphi_\delta^\dagger \partial_\nu \varphi_\lambda - (\mu \leftrightarrow \nu) \right] \label{C1}
    \\ 
  &=& -i \lambda \left[ (\partial_\mu \varphi_\lambda)^\dagger \partial_\nu \varphi_\lambda - (\mu\leftrightarrow\nu) \right].
\end{eqnarray}
This is done with the help of the identity
\begin{equation}
  \varphi_\delta^\dagger \partial_\mu \hat{n} \varphi_\lambda = (\lambda - \delta)\varphi_\delta^\dagger \partial_\mu \varphi_\lambda \label{C2} .
\end{equation}
Note that all repeated index are understood to be summed all over.
Hence the following result follows
\begin{eqnarray} \label{C1}
  Z_{\mu\nu} &=& - \epsilon^{abc}\hat{n}^a \partial_\mu \hat{n}^b \partial_\nu \hat{n}^c 
\end{eqnarray}
As a result, we reach the conclusion that the monopole term in
\begin{eqnarray}
  F_{\mu\nu}^a 
  &=& \partial_\mu B_\nu - \partial_\nu B_\mu - \epsilon^{abc}\hat{n}^a \partial_\mu \hat{n}^b \partial_\nu \hat{n}^c 
    \end{eqnarray}
$ \epsilon^{abc}\hat{n}^a \partial_\mu \hat{n}^b \partial_\nu \hat{n}^c$ 
is indeed present in the effective tensor.
In fact, the original proof is to show that the monopole structure $- \epsilon^{abc}\hat{n}^a \partial_\mu \hat{n}^b \partial_\nu \hat{n}^c $ leads right back to the above unique effective field tensor with the help of the identity Eq. \eqref{eq22} under the restriction of eigenvalue equation $\hat n \varphi_\lambda = \lambda \varphi_\lambda$ relating the normal vector $\hat n$ and scalar field $\varphi$.

This completes the proof presented in Ref. \cite{CK1} with a minor difference in notation.

\section{generalized CM type monopole solution}
If the gauge field takes the following form \cite{plb97}
\begin{eqnarray}
&&A_\mu =  f_\mu \hat n +i (f-1) \hat n \partial_\mu \hat n ~,
\\
&& Y_\mu=h_\mu -(1 - \cos \theta_1) \partial_\mu \varphi_1 ~,
\end{eqnarray}
it can shown that 
\begin{eqnarray}
\tilde \nabla_\mu \varphi_c = f \left ( {i \over 2} \partial_\mu \theta_1 + { 1 \over 2} \sin \theta_1 \partial_\mu \varphi_1  \right )~ \varphi_+
+{i \over 2} ( f_\mu-h_\mu) ~\varphi_c ~.
\end{eqnarray}
Indeed, this results follows from the fact that 
\begin{eqnarray}\label{eq38}
\partial_\mu  \varphi_c = f \left ( {i \over 2} \partial_\mu \theta_1 + { 1 \over 2} \sin \theta_1 \partial_\mu \varphi_1  \right ) ~\varphi_+
- {i \over 2} ( 1 - \cos \theta_1) \partial_\mu \varphi_1~ \varphi_c ~.
\end{eqnarray}
It is apparent that the monopole solution is derived from the partial derivative term $\partial_\mu \varphi_c$.
As a result, the gauge field components can be read off directly from the relation
\begin{eqnarray}
&& (U \tilde \nabla_\mu  U^{-1}) (U \varphi) = -{i \over 2} ( A_\mu+Y_\mu)a_-
\nonumber \\
&&
=
 f \left ( {i \over 2} \partial_\mu \theta_1 + { 1 \over 2} \sin \theta_1 \partial_\mu \varphi_1  \right ) ~a_+
- {i \over 2} ( 1 - \cos \theta_1) \partial_\mu \varphi_1~ a_- ~.
\end{eqnarray}
 by gauge transforming the scalar doublet to the unit eigenvector $a_-$ given by $U \varphi  = a_-$ with $U  =i  \exp [~ { i \over 2} \theta_1 \hat m ~] $ defined earlier.
Therefore, the result reads
\begin{eqnarray}
\left (
\begin{matrix}
A_\mu^-\\
Y_\mu-A_\mu^3a 
\end{matrix}
\right )
=
\left (
\begin{matrix}
- f \exp [-i \varphi ] \left ( {i } \partial_\mu \theta_1 +  \sin \theta_1 \partial_\mu \varphi_1  \right )  \\
- (A-B) \partial_\mu t
\end{matrix}
\right )
\end{eqnarray}
with $A^-_\mu =A^1_\mu -i A^2_\mu$.
Or equivalently,  \cite{plb97}
\begin{eqnarray}
&& A_\mu^1=-f(r)(\sin\varphi _1 \partial_\mu\theta_1
+\sin\theta_1 \cos\varphi_1  \partial_\mu\varphi_1) \\
&& A_\mu^2=f(r)(\cos\varphi_1 \partial_\mu \theta_1
-\sin\theta_1 \sin\varphi_1 \partial_\mu\varphi_1) \\
&& A_\mu^3=A(r)\partial_\mu t -(1-\cos\theta_1)\partial_\mu\varphi_1
\label{unitary}
\end{eqnarray}

Note that the electromagnetic and neutral $Z$-boson
$A_\mu^{\rm (em)}$ and $Z_\mu$ can be dressed with the Weinberg angle $\theta_{\rm w}$ as 
\begin{eqnarray}
\left( \begin{array}{cc} A_\mu^{\rm (em)} \\ Z_{\mu}
\end{array} \right)
= \left(\begin{array}{cc}
\cos\theta_{\rm w} & \sin\theta_{\rm w}\\
-\sin\theta_{\rm w} & \cos\theta_{\rm w}
\end{array} \right)
\left( \begin{array}{cc} Y_{\mu} \\ A^3_{\mu}
\end{array} \right) \nn\\
= \frac{1}{\sqrt{g^2 + g'^2}} \left(\begin{array}{cc} 
g & g' \\ -g' & g \end{array} \right)
\left( \begin{array}{cc} Y_{\mu} \\ A^3_{\mu}
\end{array} \right), 
\label{wein}
\end{eqnarray}
with the coupling constants $g, g'$ resumed explicitly.
As a result, the effective $U_1$ gauge field can be show to be
\begin{eqnarray}
&& A_{\mu}^{\rm (em)} = e\left (  \frac{1}{g^2}A(r)
+ \frac{1}{g'^2} B(r) \right ) \partial_{\mu}t 
-\frac{1}{e} \left ( 1-\cos\theta_1 \right ) \partial_{\mu} \varphi_1,   \\
&& Z_{\mu} = \frac{e}{gg'}\left (A(r)-B(r)\right ) \partial_{\mu}t,
\label{ans2}
\end{eqnarray}
that exhibits a dyon solution with a monopole structure specified by the term $-{(1/e)}(1-\cos\theta_1) \partial_{\mu} \varphi_1$.
Here $e$ represents the electric charge related to the coupling constants and Weinberg angles by
\begin{eqnarray}
e=\frac{gg'}{\sqrt{g^2+g'^2}}=g\sin\theta_{\rm w}=g'\cos\theta_{\rm w}.
\end{eqnarray}
Note again that, the existence of monopole structure is derived from the mapping of the scalar doublet with geometry $S^3$ to the $\varphi$-geometry $S^2$ with the help of the eigenvalue equation $\hat n \varphi_\pm= \pm \varphi_\pm$.
Indeed, the monopole term is derived from the $\partial \varphi_\pm$ term as shown in Eq. \eqref{eq38}.
As a result, the normalized scalar field $\varphi_\pm$ takes a special combination that leads to the existence of monopole structure.
% given by the identification of the effective $U_1$ gauge field tensor $F$ that can also be defined by
%\begin{eqnarray}
% F_{\mu\nu} && \equiv \lambda  \varphi_\lambda^\dagger G_{\mu\nu} \varphi_\lambda + i { \lambda } \; \left  [ \; ( \nabla_\mu \varphi_\lambda)^\dagger  \nabla_\nu \varphi_\lambda - \; (\nabla _\nu \varphi_\lambda)^\dagger  \nabla_\mu \varphi_\lambda\; \right ] ~.
%\end{eqnarray}
%As a result, the effective field tensor does introduce a monopole term effectively as
%\begin{eqnarray}
% F_{\mu\nu} = \partial_\mu B_\nu- \partial_\nu B_\mu  -\epsilon^{abc} \partial_\mu \hat  n^b \partial_\nu \hat n^b .
%\end{eqnarray}
Setting $\hat n = \hat r$ reduces to the CM monopole solution as expected.
We will show explicitly the relation of CM type solution with the effective gauge field tensor defined in Eq. \eqref{Fmunu}.

%\newpage
\subsection{CM monopole and covariant field tensor for monopole configuration}
Note that the CM monopole term takes the following form
\begin{eqnarray}
 A_\mu^{(em)}=\left ( 1-\cos\theta_1 \right ) \partial_{\mu} \varphi_1
 \end{eqnarray}
that represents a magnetic field of the form $B^r=-1/r^2$.
As we have mentioned earlier that the only covariant combination of gauge field tensor that behave as a monopole solution is the following effective field tensor
\begin{eqnarray}\label{FCK}
 F_{\mu\nu} && \equiv \lambda  \varphi_\lambda^\dagger G_{\mu\nu} \varphi_\lambda + i { \lambda } \; \left  [ \; ( \nabla_\mu \varphi_\lambda)^\dagger  \nabla_\nu \varphi_\lambda - \; (\nabla _\nu \varphi_\lambda)^\dagger  \nabla_\mu \varphi_\lambda\; \right ] ~
\end{eqnarray}
with $\nabla_\mu \varphi_\lambda= (\partial_\mu - iA_\mu) \varphi_\lambda = (\partial_\mu - iA^a_\mu \sigma^a/2) \varphi_\lambda$.
Here we have focus on the field for $SU_2$ only for the moment.
Indeed, we can show that the CM type solution is also derivable from the above effective tensor.
The relation follows from the identification of
\begin{eqnarray}
B^{(CM)}_\mu =i \varphi_c^\dagger \nabla_\mu \varphi_c={1 \over 2} A_\mu^3 .
\end{eqnarray}
Hence it is clear that the electromagnetic field defined for CM solution comes from the covariant combination 
$\varphi_c^\dagger \nabla_\mu \varphi_c$ with a similar covariant form.
The covariant combination is in fact the key to the consistent and covariant form of CM solution.
This is also the reason that CM solution should inevitably reduce to the original covariant effective field tensor found in Ref. \cite{CK1}.
Indeed, it can be easily shown that
\begin{eqnarray}
F^{(CM)}_{\mu \nu}= \partial_\mu B^{(CM)}_\nu -\partial_\nu B^{(CM)}_\mu 
=i [ \partial_\mu \varphi^\dagger \nabla_\nu \varphi_c+ \varphi^\dagger  \partial_\mu (\nabla_\nu \varphi_c)
- (\mu  \leftrightarrow \nu) ] ~.
\end{eqnarray}
It can hence be shown that 
\begin{eqnarray}\label{FCM}
F^{(CM)}_{\mu \nu}=  \varphi_c^\dagger G_{\mu\nu} \varphi_c + i  \; \left  [ \; ( \nabla_\mu \varphi_c)^\dagger  \nabla_\nu \varphi_c - \; (\nabla _\nu \varphi_c)^\dagger  \nabla_\mu \varphi_c\; \right ] ~.
\end{eqnarray}
Consequently,  $F^{(CM)}_{\mu \nu}$ agrees with the effective covariant field tensor defined in Eq. \eqref{FCK}.
Note also that $\varphi_+$ and $\varphi_c=i \varphi_-$ are conjugate to each other under the dyon charge conjugation transformation.

\subsection{special solution to the monopole}
Given the model with action \eqref{action}, the equations of motion can be shown to be
\begin{eqnarray} \label{eq68}
&& \nabla_\mu G^{\mu \nu} 
= 
{i \over 2} \chi^2 (\nabla^\nu  \varphi \varphi^ \dagger - c.c.) ,
\\ && \label{eq69}
\partial_\mu Y^{\mu \nu} = 
{i \over 2} \chi^2 (\varphi^ \dagger \nabla^\nu \varphi - c.c.) ,
\\ &&
(1 - \varphi \varphi^\dagger ) \nabla_\mu ( \chi^2 \nabla^\mu  \varphi)=0,
\\
&&
\partial^2 \chi- \chi \nabla^\nu  \varphi ^\dagger \nabla^\nu  \varphi = {\lambda \over 2} (\chi^2-v^2)\chi .
\end{eqnarray}
Hence the field equations reduces to 
\begin{eqnarray}
&& \chi'' +\frac{2}{r} \chi' -(l^2+p^2) \frac{f^2}{4r^2}\chi
=-\frac{1}{4}(A-B)^2\chi +\frac {\lambda}{2}\big(\chi^2
-v^2 \big)\chi ,
\label{eqchi} 
\\ &&
{f}''-l^2 \frac{f^2-1}{r^2}f=\big(\frac{g^2}{4}\chi^2
- A^2\big)f,
\label{eqf} 
\\ &&
{A''}+\frac{2}{r} {A'}-(l^2+p^2)\frac{f^2}{r^2}A
=\frac{g^2}{4}\chi^2(A-B), 
\label{eqA}
\\ &&
{B''} +\frac{2}{r} {B'}
=-\frac{g'^2}{4} \chi^2 (A-B)
\label{eqB}  
\end{eqnarray}
for the ansatz $\varphi=\varphi_c$ obeying the eigenvalue equation $\hat n \varphi_c=-\varphi_c$.
Note that Eqs. \eqref{eqA} and \eqref{eqf}  follow from the component equations $\nabla^\mu G_{\mu 0}$ and $\nabla^\mu G_{\mu \theta}$ of Eq. \eqref{eq68} respectively.
In addition, the Eq. \eqref{eqB} follows from the component equation $\partial^\mu Y_{\mu 0}$ of Eq. \eqref{eq69}.
We have also used the hermitian matrix identities $A_0=A \hat n$, $A_\theta=(f-1) \hat \varphi_1/r$ and $A_{\varphi}=-(f-1) \hat \theta_1/r$ associated with the ansatz $\varphi_c(\hat n)$.

We would like to focus on the effect of $l$ by setting $p=0$ as a simple demonstration.
In addition, we will also focus on a special solution with the property that $\varphi^ \dagger_c \nabla^\nu  \varphi_c=0$.
As a result, the constraint leads to the solution with $A=B$.
Moreover, the monopole term takes the following form in this case
\begin{eqnarray}
A_\mu =- (1-\cos \theta) l \partial_\mu \varphi~.
\end{eqnarray}
It hence represents a monopole solution with magnetic charge $l$ times the CM solution.

In addition, Eqs. \eqref{eqA} and \eqref{eqB} also imply that $f=0$ too.
Consequently,  the scalar field $\chi$ decouples from the all other fields in this model.
Hence, the solution for $A$ an $B$ can be integrated directly as
\begin{eqnarray}
A=B=A(r_0)- A'(r_0) {r_0^2 \over r^2} 
\end{eqnarray}
with a reference point set at $r=r_0$.
Note that the solution with $f=0$ turns off effectively all the effects of $p$ and $l$ from the field equations. 
This is indeed the most simple set of solution for the system.

\subsection{Dirac monopole and quantization of magnetic charge}
The spatial part of the gauge field tensor derived from the magnetic term in the CM gauge field solution
\begin{eqnarray}
A^3_\mu =-{ 1 \over e} (1-\cos \theta)\partial_\mu \varphi
\end{eqnarray}
is
\begin{eqnarray}
F_{ij}=-{ 1 \over e} \sin \theta \partial_i \theta \partial_j \varphi
\end{eqnarray}
As a result,  $B^i=\epsilon^{ijk}F_{jk}/2= -\epsilon^{ijk} \sin \theta \partial_j \theta \partial_k \varphi_1/e$.
Note that the totally skew-symmetric Levi-Civita 3-tensor is defined as $\epsilon^{ijk}=e^{ijk} /\sqrt{g}$ with $g= \det g_{ij}$ and $e^{ijk}=1$ for cyclic permutation of $i, j, k$. 
Therefore, the only non-vanishing magnetic field in spherical coordinate is given by
\begin{eqnarray}
B^r=\epsilon^{r \theta\varphi} F_{\theta \varphi}/2=-{ 1 \over er^2} \partial_\theta \theta \partial_\varphi \varphi =-{ 1 \over er^2} .
\end{eqnarray}
Hence $B^r=-{ 1 / (er^2)}$ for  the ($\hat n =\hat r$) CM solution.

%%%

On the other hand, the gauge field for the generalized CM gauge monopole solution is
\begin{eqnarray}
A_\mu =-{ 1 \over e} (1-\cos \theta_1)\partial_\mu \varphi_1~.
\end{eqnarray}
%is
%\begin{eqnarray}
%F_{ij}=-{ 1 \over e} \sin \theta_1 \partial_i \theta_1 \partial_j \varphi_1
%\end{eqnarray}
We can transform the coordinate $\hat r$  to the generalized coordinate specified by $\hat n$.
In the new coordinate $n, \theta_1, \varphi_1$ the non-vanishing field  tensor is
\begin{eqnarray}
\tilde F_{\theta \varphi}=-{ 1 \over e} {\sin \theta_1 } (\partial_{\theta_1} \theta_1 \partial_{\varphi_1} \varphi_1-
\partial_{\varphi_1} \theta_1 \partial_{\theta_1} \varphi_1)= -{ 1 \over e} {\sin \theta_1 }  ~.
\end{eqnarray}
Therefore, the only non-vanishing magnetic field in spherical coordinate is also given by
\begin{eqnarray}
\tilde B^n=-{ 1 \over er^2}  
\end{eqnarray}
in the new coordinate.
In addition, we can show that the 3-space tensor 2-form $F =-{ (1 / e)} \epsilon_{abc}\hat n^a d \hat n^b d \hat n^c$ with monopole term  %$F= -{ (1 / e)} \epsilon_{abc}\hat n^a d \hat n^b d \hat n^c$
derives the winding number for the mapping from $\hat n$-sphere to the $\hat r$-sphere when it is integrated as 
\begin{eqnarray}
\int F = -{ 1 \over e} \int  \epsilon_{abc}\hat n^a d \hat n^b d \hat n^c = -{ k \over e}
\end{eqnarray}
for some integer $k$.
Here gauge field one-form is defined as $B=B_i dx^i$, the tensor 2-form is defined as $F=F_{ij}dx^idx^j$.
As a result, it is known that the generalized CM monopole solution specified by $\hat n$ maps the $\hat n$-sphere to the $\hat r$-sphere by wrapping around the $\hat r$-sphere $k$ times.

Note also that the magnetic field for a Dirac monopole with magnetic charge $g$ (Appendix A) is known to be
$B^r=-{g / (4 \pi r^2)}$ with a quantization rule $ eg=2n \pi$ derived from the Aharonov–Bohm (AB) phase factor
\begin{eqnarray}
\exp \left  [~ {ie \oint A \cdot dx}~ \right ]=\exp \left  [~ {ie \int B \cdot dS}~ \right ]=\exp[~ i 2n\pi~] ~.
\end{eqnarray}
The AB factor has to be uniquely defined in quantum mechanics.
Hence the quantization rule for the CM monopole takes the form $ge=4\pi$.
On the other hand, the quantization rule for the generalized CM monopole takes the form $ge=4pl\pi$ if $\theta_1=p \theta$ and $\varphi_1=l \varphi$ as a simple generalization of the CM monopole.
Note that $p, l$ should be both integers such that the mapping of $\hat n$-sphere can wrap around the $\hat r$-sphere $pl$ times completely. 
Accordingly, the AB phase factor can also be uniquely defined.
In addition, the angle $\varphi_1$ will rotate $l$ times when the azimuthal angle $\varphi$ rotates $2\pi$ for $\varphi_1=l \varphi$.
As we have shown earlier that the eigenvector $\varphi_+$ is a gauge transform of base vector $a_\pm$ via the relation $\varphi_+= iU^{-1} a_+$ with $iU^{-1}=\exp \left  [~ -{i \over 2} \theta_1 \hat m ~ \right ]$ and $\hat m= (\sin \varphi_1, - \cos \varphi_1,0) $.
Hence the unit vector $\hat m$ will also rotate $l$ times when the azimuthal angle $\varphi$ rotates $2\pi$. 

\section{conclusion}

A new type of non-abelian gauge field dyon solutions known as Cho-Maison (CM) monopole are found in Ref. \cite{plb97,yang}.
An  $SU_2$  scalar field ansatz was proposed to take the form 
$\varphi_c^t=i( \cos (\theta/2)  \exp(-i \varphi), - \sin (\theta/2)~)$. 
We have pointed out that the scalar doublet is in fact a special solution to the eigenvalue equation $\hat n \varphi_c \equiv  \hat n^i \cdot  \sigma^i  \varphi_c = -  \varphi_c$ with $\hat n =\hat r$. \cite{CK1}
The unit vector $\hat n =(\sin \theta_1 \cos \varphi_1, \sin \theta_1 \sin \varphi_1, \cos \theta_1)$ with $\theta_1(x), \varphi_1(x)$ denotes in fact the local spherical angles (functions of space time) associated with the unit vector $\hat n(x)$.
As a result, a new set of monopole solutions are generated accordingly.

In particular, the existence of monopole solution has to do with the identification of a special combination of gauge field to exhibit the monopole structure with
\begin{eqnarray}\label{4.28}
  F_{\mu\nu} 
&&  = \partial_\mu B_\nu - \partial_\nu B_\mu - \epsilon^{abc}n^a \partial_\mu n^b \partial_\nu n^c 
\end{eqnarray}
The monopole term $\epsilon^{abc}n^a \partial_\mu n^b \partial_\nu n^c $ can be shown to be uniquely given by the identification of gauge field tensor 
\begin{eqnarray}
 F_{\mu\nu} && \equiv \lambda  \varphi_\lambda^\dagger G_{\mu\nu} \varphi_\lambda + i { \lambda  } \; \left  [ \; ( \nabla_\mu \varphi_\lambda)^\dagger  \nabla_\nu \varphi_\lambda - \; (\nabla _\nu \varphi_\lambda)^\dagger  \nabla_\mu \varphi_\lambda\; \right ] ~
\end{eqnarray}
with the help of the eigenvalue equation $\hat n \varphi_\pm = \pm \varphi_\pm$.

The existence of gauge monopole solution has to do with the identification of gauge field $\phi$ with the eigenvector to the eigenvalue equation $\hat n^i T^i \varphi_\lambda= \lambda \varphi_\lambda$, or equivalently, \cite{CK1}
\begin{eqnarray}
\hat n  \varphi_\lambda= \lambda \varphi_\lambda 
\end{eqnarray}
In addition, we have also shown that the scalar field $\phi$ can be decomposed as $\phi=\chi \varphi_0 \varphi_\lambda$.
$\chi$ and $\varphi_0= \exp [~i \psi~]$ represent the norm and abelian phase factor of $\phi$ respectively.
We have also shown explicitly that the CM monopole can be generalized straightforwardly to a new set of monopole solutions by extending $\hat r \to \hat n$.
It is also shown that the $F^{(CM)}_{\mu \nu}$ defined by the CM solution agrees with the effective covariant field tensor defined in Eq. \eqref{FCK}.

A brief outline of Dirac and 't Hooft-Polyakov monopole is also presented in appendix for reference.
The main result of  Ref. \cite{CK1} in matrix formulation was also outlined briefly in this paper for reference.
The materials shown in this paper will hopefully be helpful in generalizing many known monopole solutions in a more systematic way.

{\bf Acknowlegement } This paper is supported in part by the Ministry of Science and Technology (MOST) of
Taiwan under Contract No. MOST 109-2112-M-009-001.
\appendix

\section{Dirac string}\label{sec:Dirac string}
Dirac proposed the existence of monopole obeying the equation
\begin{eqnarray}\label{point_magnetic_charge}
  \nabla\cdot\vec{B} &=& g\delta(\vec{r}) .
\end{eqnarray}
Equivalently, the monopole charge is given by
\begin{eqnarray}\label{point_magnetic_charge_form_B_field}
  g &=& \int_{S}\vec{B}\cdot d\vec{S}
\end{eqnarray}
with $S$ the closed surface enclosed the origin.

It is known that $g = 0$ if $\vec{B} = \nabla\times\vec{A}$.
Hence  $g \neq 0$ implies the existence of singularity in $\vec{A}$ on the surface of $S$.
This implies immediately the existence of a singular string.
Indeed, a string sitting in the negative z-direction is shown below for a simple demonstration.

\tikzstyle{spring}=[thick,decorate,decoration={zigzag,pre length=0.1cm,post
  length=0.1cm,segment length=4}]
\begin{eqnarray}\label{graph_1}
\raisebox{5pt}{
\begin{tikzpicture}
  \draw [very thick] (0,0) circle (1.5pt) ;
  \draw [->,thick] (-3pt,0) -- (-50pt,0) ;
  \draw [->,thick] (3pt,0) -- (50pt,0) ;
  \draw [->,thick] (0,-3pt) -- (0,-50pt) ;
  \draw [->,thick] (0,3pt) -- (0,50pt) ;
  \draw [->,thick] (-3pt,-3pt) -- (-35pt,-35pt) ;
  \draw [->,thick] (-3pt,+3pt) -- (-35pt,+35pt) ;
  \draw [->,thick] (3pt,3pt) -- (35pt,35pt) ;
  \draw [->,thick] (3pt,-3pt) -- (35pt,-35pt) ;
\end{tikzpicture}}
 \quad &\raisebox{52pt}{=} \quad
\begin{tikzpicture}
  \draw [very thick] (0,0) circle (1.5pt) ;
  \draw [->,thick] (-3pt,0) -- (-50pt,0) ;
  \draw [->,thick] (3pt,0) -- (50pt,0) ;
  \draw [thick] (-1pt,-55pt) -- (-1pt,-1pt) -- (1pt,-1pt) -- (1pt,-55pt) ;
  \draw [->,very thick] (0,-32pt) -- (0,-19pt) ;
  \draw [->,thick] (0,3pt) -- (0,50pt) ;
  \draw [->,thick] (-3pt,-3pt) -- (-35pt,-35pt) ;
  \draw [->,thick] (-3pt,+3pt) -- (-35pt,+35pt) ;
  \draw [->,thick] (3pt,3pt) -- (35pt,35pt) ;
  \draw [->,thick] (3pt,-3pt) -- (35pt,-35pt) ;
   \draw[spring] (0,-0.65) -- (0,-0.1);
  \draw[spring] (0,-1.9) -- (0,-0.8);
\end{tikzpicture}
  \quad \raisebox{52pt}{+} \qquad
\begin{tikzpicture}
  \draw [very thick] (0,0) circle (1.5pt) ;
  \draw[spring] (0,-1.0) -- (0,-0.1);
  \draw[spring] (0,-1.9) -- (0,-1.2);
  \draw [thick] (-1pt,-55pt) -- (-1pt,-1pt) -- (1pt,-1pt) -- (1pt,-55pt) ;
  \draw [<-,very thick] (0,-32pt) -- (0,-19pt) ;
\end{tikzpicture}
 \qquad \nonumber
\end{eqnarray}
\centerline{Fig. 1\;\; Dirac string and monopole}

Magnetic field derived from Dirac's monopole shown on the left half of the figure can be decomposed as a combination of a solenoid field  and a string induced field on the right half of the figure.
Indeed, the solenoid generates a magnetic field of the form
\begin{eqnarray}\label{B_solenoid}
  \vec{B}_{ \text{sol} } &=& \frac{g}{4\pi r^{2}}\hat{r} + g\theta(-z)\delta(x)\delta(y)\hat{z}
\end{eqnarray}
such that $\nabla\cdot\vec{B}_{ \text{sol} } = 0$.
As a result, it can be shown that
\begin{eqnarray}\label{B_solenoid_in_curl_A}
  \vec{B}_{ \text{sol} } &=& \nabla\times\vec{A} \\
  \vec{B} &=& \frac{g}{4\pi r^{2}}\hat{r} 
  = \nabla\times\vec{A} - g\theta(-z)\delta(x)\delta(y)\hat{z}
\end{eqnarray}
It can also be shown that
\begin{eqnarray}\label{string_vector_potential_A}
  \vec{A} &=& \frac{g}{4\pi r}\frac{(1-\cos\theta)}{\sin\theta}\hat{\varphi}
\end{eqnarray}

\section{$SO_3$ regular monopole}

In 1974, ’t Hooft and Polyakev \cite{10} found a regular monopole solution in $SO_3$ Georgi-Glashow Model given by the Lagrangian
\begin{eqnarray}\label{Lagrangian}
  \mathcal{L} &=& -\frac{1}{4}G_{\mu\nu}^{\ \ a}G^{\mu\nu a} + \frac{1}{2}D^{\mu}  \vec{\phi} \cdot D_{\mu}\vec{\phi} - V(\vec{\phi}) \nonumber \\
  \text{with\qquad} G_{\mu\nu}^{\ \ a} &=& \partial_{\mu}A_{\nu}^{\ a} - \partial_{\nu}A_{\mu}^{\ a} - e\epsilon^{abc}A_{\mu}^{\ b}A_{\nu}^{\ c}\nonumber \\
  D_{\mu}\vec{\phi} &=& \partial_{\mu}\vec{\phi} - e\vec{A}_{\mu}\times\vec{\phi} \nonumber \\
  V(\vec{\phi}) &=& \frac{\lambda}{4}(\vec{\phi}^{2} - a^{2})^{2}
\end{eqnarray}
The equation of motion can be shown to be
\begin{eqnarray}\label{least_action_principle}
  D_{\mu}\vec{G}^{\mu\nu} &=& e\vec{\phi}\times D^{\nu}\vec{\phi} \nonumber \\
  D^{\mu}D_{\mu}\vec{\phi} &=& -\lambda\vec{\phi}\times(\vec{\phi}^{2} - a^{2})\end{eqnarray}
In addition the dual field tensor $ {^*}{\vec{G}}{_{\mu\nu}} = \epsilon^{\mu\nu\rho\sigma}\vec{G}^{\rho\sigma}/2$ satisfies the equation
\begin{eqnarray}\label{least_action_principle1}
  D_{\mu} {^*}{\vec{G}}{_{\mu\nu}} &=& 0 \text{\quad,with \qquad}
\end{eqnarray}
't Hooft proposed a combination of gauge field tensor to form an effective $U_1$ field tensor  $F_{\mu\nu}$ of the following form
\begin{eqnarray}\label{Hooft_Tensor}
  F_{\mu\nu} &=& \frac{1}{{{\phi}}}\vec{\phi}\cdot\vec{G}_{\mu\nu} - \frac{1}{ e{{\phi}}^{3} } \vec {\phi}\cdot\left[ (D_{\mu}\vec{\phi})\times(D_{\nu}\vec{\phi}) \right] \;  ,
\end{eqnarray}
with $\phi \equiv |\vec \phi|$ \cite{7,8,13} .
\begin{eqnarray}
   \qquad B^{k} = -\frac{1}{e}\frac{r^{k}}{r^{3}}, \qquad F_{ij} = -\epsilon_{ijk}B^{i} 
\end{eqnarray}
%\centerline{ \text{當} $ r \to \infty $\text{時}。}
Indeed, a scalar field ansatz is proposed as follows
\begin{eqnarray} \label{static_ansatz} \label{SO3phi}
  \phi^{a} &=& \frac{r^{a}}{er^{2}} H(x) \nonumber\\
  A_{a}^{i} &=& - \epsilon_{aij}\frac{r_{j}}{er^{2}} \left[ 1-K(x) \right]  \nonumber\\
  A_{a}^{0} &=& 0
\end{eqnarray}
with $ x = aer $, $H$ and $K$ are functions to be determined by the field equations.
As a result, the monopole solution gives a regular magnetic field of the form
\begin{eqnarray}
  -\epsilon_{ijk}B^{k} &=& \frac{\vec{\phi} \cdot \vec{G}_{ij}}{{\phi}} = \epsilon_{ijk} \frac{r^{k}}{er^{3}} \nonumber \\
  B^{k} &=& -\frac{r^{k}}{er^{3}} .
\end{eqnarray}
Note that the normalized scalar field $\hat \varphi=\hat r$ in Eq. \eqref{SO3phi} is indeed an eigen solution to the eigen equation  $\hat r \cdot T\hat  \varphi=0$ with eigenvalue $0$.

\end{CJK}
\end{document}